\documentclass[twocolumn,showpacs,preprintnumbers,amsmath,amssymb]{revtex4}
\usepackage{graphicx}
\usepackage{dcolumn}
\usepackage{bm}

\begin{document}

\title{Properties of high-T$_C$ copper oxides from the nearly-free electron
model.}

\author{T. Jarlborg}

\affiliation{
DPMC, University of Geneva, 24 Quai Ernest-Ansermet, CH-1211 Geneva 4,
Switzerland
}

%\date{\today}

\begin{abstract}

The generic band structure of high-T$_C$ copper oxides is simulated by
the nearly free-electron model (NFE) in two dimensions (2-D) with parameters from
band calculations. Interaction between phonons and spin waves will cause
potential modulations and pseudogaps, and the strength of the  
modulations, the wave lengths and the doping, are all related. 
A Fermi-surface "arc" is found for dynamic spin/phonon
waves.  The confinement of superconductivity
between two limiting dopings can be a result of competition with the pseudogap
at low doping and weak coupling at high doping.

\end{abstract}


\maketitle

%\section{\label{sec:level1}Introduction}

An important structural feature of high-T$_C$ materials
is the stacking of almost 2-dimensional (2-D) CuO planes. All
high-T$_C$ cuprates have at least one of these layers in the unit
cell, and they make the band structure fairly simple with
one or more $M$ centered Fermi surface (FS) cylinders. 
Band calculations and photoemission agree essentially on
this fact, except for details and for undoped materials, which
often are antiferromagnetic (AFM) insulators \cite{dama}.  
The density of states (DOS) at the Fermi energy ($E_F$) is mainly of Cu-d character 
with some O-p admixture.
The views on the mechanism
for superconductivity and normal state properties diverge.
Extensive experimental works show that not only the high T$_C$, 
but also many normal state properties, are unusual.
These include pseudogaps and magnetic fluctuations, both with
rich dependencies as function of doping, $x$, and temperature, T.
The importance of phonons is evidenced by isotope effects on $T_C$ and
pseudogaps \cite{khas,gwe}, and magnetic fluctuations are detected
by neutron scattering \cite{tran}.
 
Here is presented a NFE model for high-T$_C$
properties with use of parameters coming from ab-initio
band calculations for pure and hole doped high-T$_C$ materials \cite{tj4,tj5}.
These band calculations show important coupling between phonon and spin waves.
The difficulty with ab-initio calculations is that a unit cell with interesting
phonon displacements and/or spin waves needs to be very large.
In addition, the bands are not obtained within the normal
Brillouin Zone (BZ), but within the down-folded zone.
Such calculations are so far limited to spin waves and phonons
along one direction only, while in reality one can expect
modulations along $\hat{x}$ $\it and$ $\hat{y}$ (checker board
modulations rather than stripes). The qualitative
results from the one-dimensional (1-D) NFE model and ab-initio calculations
are the same, and it is worthwhile to
extend the NFE-model to 2-D.

A periodic potential perturbation, 
$V(\bar{x}) = V_Q exp(-i\bar{Q} \cdot \bar{x})$,
will open a gap of size 2$V_Q$ at the zone boundary, $\bar{k} = \bar{Q}$/2, in the 1-D NFE bands \cite{zim,tj1}. 
This well-known NFE result can be applied to the AFM spin arrangement on neighboring Cu along [1,0,0]
(with wave vector $Q$)
in undoped insulating cuprates. 
An additional modulation, with wave vector $\bar{q} < \bar{Q}$, modifies the potential,
$V(\bar{x}) = 
V_Q exp(-i\bar{Q} \cdot \bar{x}) exp(i\bar{q} \cdot \bar{x})$.
% = V_Q exp(-i(\bar{Q}-\bar{q}) \cdot \bar{x})$. 
The gap moves away from $\bar{Q}/2$ to ($\bar{Q}-\bar{q}$)/2, as for
ab-initio bands in supercells of different lengths.
 The combined potential modulation leads to stripe like phonon or spin-wave patterns, 
with short wave lengths
(periodicity, with wave vector $\vec{q}$) at high doping and long ones when $x \rightarrow 0$, as shown in previous
band calculations \cite{tj5}.

%\section{Results}

\begin{figure}
%\vskip -5mm
\includegraphics[height=7.0cm,width=8.0cm]{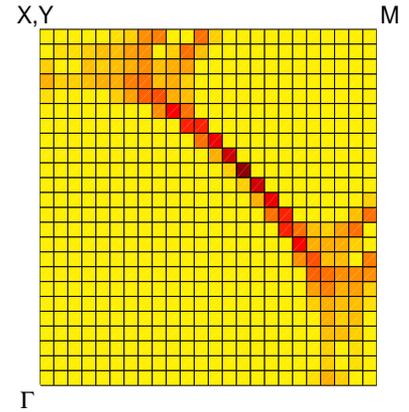}
\caption{Calculated 2-D NFE Fermi surface for a strong dynamic
spin-phonon fluctuation with $\vec{q}$=(0.85,0.90). 
}
\label{fig1}
\end{figure}

Simultaneous modulations along $\hat{x}$ and $\hat{y}$ are yet too difficult
for an ab-initio band approach, at least for realistic wave lengths. 
%In 2-D it is not expected that the gap in the total DOS will be fixed at the same energy as in
%the case of a 1-D wave, because  $V_Q$ may affect the band differently at different k-points. 
An extension of the NFE model for
potential modulations along $\hat{x}$ and $\hat{y}$ leads to a 3x3 eigenvalue problem of the form
$E-k_x^2-k_y^2$, $E-(k_x-Q_x)^2-k_y^2$, and $E-k_x^2-(k_y-Q_y)^2$ in the diagonal, and
$V_Q$ as non-diagonal terms. 
The G-vectors 0, $\bar{Q}_x = \bar{Q}-\bar{q_x}$ and
$\bar{Q}_y = \bar{Q}-\bar{q_y}$ are considered in the basis
with $\bar{q_x}$ and $\bar{q_y}$ along $\hat{x}$ and $\hat{y}$. The bands are represented
in the reduced zone as for the normal unit cell. 
The band width from the $\Gamma$-point
up to $\hat{k}_F$ is about 0.125 Ry when $V_Q, q_x, q_y$ and $x$
are all zero
and the effective mass is 2.5. The position
of the van Hove singularity (when the bands touch $X$ and $Y$) corresponds to
$x \approx 0.2$. Potential modulations will open gaps
near $X$ and $Y$,
while not much happens between $\Gamma$ and $M$  \cite{tj5}.

Superconductivity and pseudogaps appear typically at 100K, and we will estimate the $V_Q$-parameters for
this temperature. The mean value of atomic displacements $u = \sqrt(3 k_B T/ K)$, where the force constant ($K$) for
O vibrations is of the order 25 $eV/\AA^2$ for $\hbar \omega \approx$ 50 meV, is close to
0.01$a_0$, where $a_0$ is the lattice constant. Zero-point motion is just a bit larger, but it is not selective to one phonon \cite{grim}.
The mean value of magnetic moment fluctuations, $m_0$, is obtained through $k_B T = \frac{1}{2} F m^2$, where $F = d^2E/dm^2$
and $E$ is the total energy. For a short spin-phonon wave in HgBa$_2$CuO$_4$ \cite{tj4}, $V_Q^{sf}$ is about 6 mRy for $m_0 \sim 0.09 \mu_B$
per Cu.

%\subsection{Nearly free electron models.}

\begin{table}[b]
\caption{\label{table1}Spin density coefficients (A cos$^2(\pi x/2)$ and B sin$^2(\pi x/2)$) for the NFE state 
below the
gap at the zone boundary along $q_x$. The exchange splitting for spin fluctuations, $V_Q^{sf}$, is
estimated from the scaling procedure as described in the text.
%The $V_Q^{sf}$ in the last two lines are reduced additionally, because of the the low $T^*$.
The last column shows the potential parameter for phonons, $V_Q^{ph}$, which are interpolated from band results
for $Q \leq 0.9$. A saturation is assumed for $Q \geq$ 0.9. }
\vskip 5mm
\begin{center}
\begin{tabular}{|l|c|c|c|c| }
\hline
$Q_x$ & $Q_y$ & (A-B) & $V_Q^{sf}$ (mRy) & $V_Q^{ph}$ (mRy)\\
\hline \hline
0.99 & 0.99 &  0.98 & 32 & 12 \\
0.95 & 0.95 &  0.70 & 22 & 12 \\
0.90 & 0.90 &  0.47 & 15 & 11 \\
0.83 & 0.83 &  0.33 & 11 & 9 \\
0.75 & 0.75 &  0.25 &  8 & 9 \\
\hline
0.50 & 0.50 &  0.18 &  6 & 7 \\
\hline
0.75 & 0.85 &  0.25 &  7 & 9 \\
0.75 & 0.95 &  0.25 &  6 & 9 \\
\hline
\end{tabular}
\end{center}
\end{table}

The origin of $V_{Q}$ is two-fold. Structural distortions of phononic origin contribute 
equally to the two spin components of $V(\bar{x})$, while for a spin wave there is an opposite phase
of the two spin potentials. A static spin-polarized modulation of the potential determines AFM order in
the undoped case. Phonons, and probably also long-wave spin-modulations, are dynamic. From the 
self-consistent band calculations
for 'half breathing' phonons along [1,0,0] it is estimated that $V_Q^{ph}$ varies from 
5 to 11 mRy when the wave length varies from 4$a_0$ to 16$a_0$. This is when the atomic
displacement is of the order 0.01$a_0$, as when T $\approx$ 100K \cite{tj5}.
The phonon amplifies
a spin wave, which has twice a long wave length 
as the phonon, but the two types of waves enforce a common gap at the same energy \cite{tj2}.

The self-consistent convergence of long-wave spin configurations is very slow in band calculations.
In order to estimate the spin-polarized part $V_Q^{sf}$ we rely
on the band results for a short spin wave and do NFE-scaling for longer waves. Band calculations on
undoped Hg$_2$Ba$_4$Cu$_2$O$_8$ show that an AFM, zero-gap state can be stabilized with a spin-splitting
of 23 mRy on Cu \cite{tj4}. (Larger gaps are possible when using a different density functional in 
band calculations for La$_2$CuO$_4$ \cite{per}).
This determines the $V_Q^{sf}$ for the basic AFM spin arrangement between nearest Cu neighbors.
A well-known feature of the 1-D NFE model is the spacial separation of the states below the gap ($\Psi^2_-(\hat{x})
= sin^2(Gx/2))$ and above the gap ($\Psi^2_+(\hat{x})
= cos^2(Gx/2))$ \cite{zim}. Let the state below (above) the gap coincide with the Cu with the attractive 
(repulsive) potential for one 
spin.  The spin density of the first (second) spin is given by
the $sin^2$- ($cos^2$)-term, and
the phase on the nearest Cu neighbor differ by $\pi$.
 The densities feed the exchange splitting of the potentials. If they diminish it leads to
a smaller exchange, which leads to smaller densities and so on, in a self-consistent manner.
Thus, the spin density is a
driving force behind the near neighbor AFM configuration, and it depends on the optimal occupation of the two states.

An additional modulation will reduce the spacial separation
of the two states at $G/2$. Table I shows how the lower state is mixed as function of
$\vec{Q_x}$. As $V_Q^{sf}$ depends on the near-neighbor interaction one can expect a reduction of $V_Q^{sf}$
by the coefficients given in Table I.
This does not include self-consistent feedback, or the closing of the pseudogap at $T^*$ (see later), effects
which both should decrease $V_Q^{sf}$.  Usually 
$T^*$ is well above the 100K at which the parameters are estimated. But when $T^*$ goes to zero
at large doping, it will reduce $V_Q^{sf}$ more (the two last lines in Table I
include rather arbitrary reduction factors of 0.9 and 0.8).
On the other hand, the coefficients $V_Q^{sf}$ should increase because
of coupling to phonons for all $\vec{q} >$ 0. This effect is estimated to increase the moments
by at least 30 percent for a short phonon wave with appropriate value of $u$
\cite{tj3}, and the values in Table I include a factor of 1.3. 
It can be recalled that $V_Q^{sf}$ at $q_x = q_y = 0.5$ from the simple rescaling, 6 mRy, agree with the independent estimate
from the best converged band calculation for that spin wave (4$a_0$) in doped HgBa$_{2}$CuO$_4$ \cite{tj4}.

\begin{figure}
%\vskip -5mm
\includegraphics[height=6.0cm,width=8.0cm]{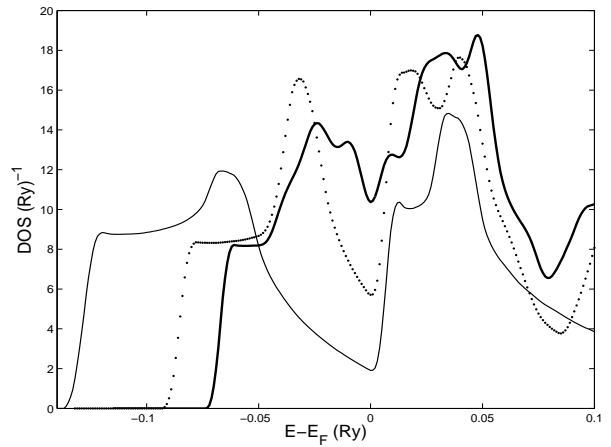}
\caption{Calculated DOS for $x$=0.02 (thin line), $x$=0.16 (dotted)
and $x$=0.26 (heavy line). The latter case is for non-equivalent modulations
along $\hat{x}$ and $\hat{y}$, and a secondary dip below $E_F$ is seen.
}
\label{fig2}
\end{figure}

\begin{figure}
%\vskip -5mm
\includegraphics[height=7.0cm,width=8.0cm]{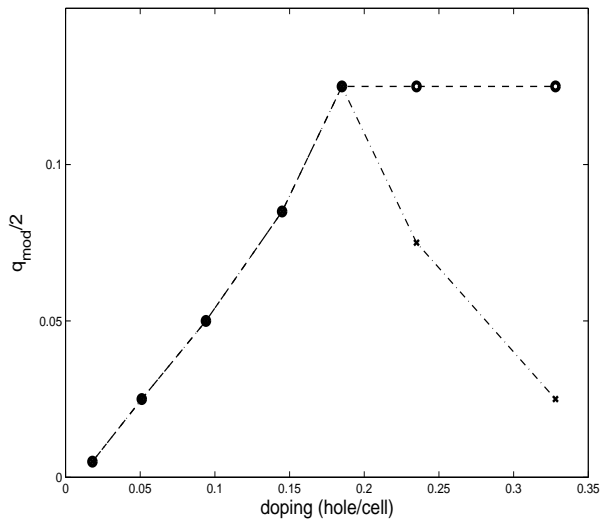}
\caption{The relation between doping $x$ and q-vectors of the
modulation. The q-vectors along $\hat{x}$ and $\hat{y}$ are equal for doping below $\sim$
0.18. The modulation along one direction remains fixed (8 $a_0$ for the spin part)
for larger doping, 
with a weaker modulation of longer periodicity in the perpendicular direction.
}
\label{fig3}
\end{figure}

\begin{figure}
%\vskip -5mm
\includegraphics[height=7.0cm,width=8.0cm]{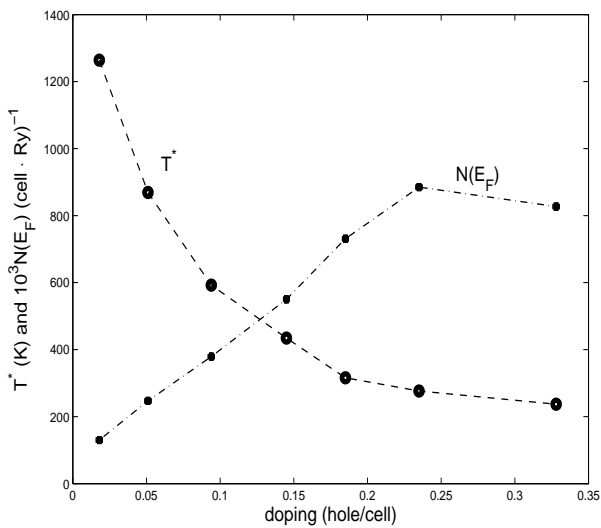}
\caption{The DOS at $E_F$ and $T^*$ as function of doping $x$. The DOS is at the minimum of the 
pseudogap and $T^*$ is 1/4th of $V_Q^{sf}$.
}
\label{fig4}
\end{figure}

An example of the FS is shown in figure 1, which is a sum of several calculations with
$V_Q$ ranging from 10 to 30 mRy in order to simulate dynamical waves. 
The position of $E_F$ is chosen at the minimum of the sum of the
partial DOS functions, and the plot is finally symmetrized along the $\Gamma$-$M$ line.
 The FS remains as a sharp "arc" on
the diagonal direction, while it is washed out near the limits of the zone because of the fluctuations of $V_Q$.
The result is compatible with the observation of a small section of a FS-arc at low T \cite{nor}. The arc
widens for larger T, when most of the potential modulations and the pseudogap are gone.
Static potential modulations will bend the outer sections of the FS towards the $X,Y-M$
lines \cite{tj5}, which in a repeated zone looks like a second "ghostband".

Calculations of the DOS (fig. 2) and the relation between doping and q-vectors 
(fig. 3) are made using $V_Q = V_Q^{sf}+V_Q^{ph}$ from Table I as input. The
Q-vectors are allowed to vary from 1.0 to 0.75 in $\hat{x}$ and $\hat{y}$ directions, and the doping is optimal
when $E_F$ coincides with the energy at the DOS minimum, at the pseudogap. Three examples of doping
are shown in figure 2. At low doping ($x \leq 0.18$) there is
a nearly linear relation between $x$ and $q$ (see fig. 3), and the Q-vectors along $\hat{x}$ and $\hat{y}$
have equal lengths down to the minimum at 0.75 of the zone boundary limit.  
There is no possiblity to obtain a minimum in the DOS for larger doping with equal 
$\mid \vec{Q}_x \mid$ and $\mid \vec{Q}_y \mid$,
the values of $V_Q$ are too small. However, if one of the Q-vectors remains
fixed at the value 0.75 (assumed to correspond to the shortest possible magnetic modulation \cite{tran}) while the other one increases,
it is possible to follow the pseudogap further towards large $x$, with $V_Q$ coming from Table I. 
(The larger of the two Q-vectors makes a dip in the DOS, but it is weak and below $E_F$, see fig. 2, so its modulation should be harder to detect.)  
The result in Fig. 3 is qualitatively similar to the doping dependence observed in 
La$_{(2-x)}$Sr$_x$CuO$_4$ by Yamada et. al. \cite{yam}, although
saturation of the periodicity occurs near $x \sim 0.18$ compared to about 0.12 as observed.
Other combinations of $Q$-vectors and $V_Q$ can give larger $x$, but with weaker gaps
or unrealistic values of $V_Q$.

The unperturbed NFE band is perfectly isotropic, but the real band structure
may have different dispersion in different directions. A FS which extends more towards the diagonal
than towards $X$ and $Y$, can be modelled by an anisotropic effective mass. Here, through
multiplication of the mass with 
(($\mid k_x\mid+\mid k_y\mid)$/$(\mid k_x+k_y\mid$))$^{\frac{1}{3}}$ we test a $\sim$ 10 percent
anisotropy. The
result is that the scale of the doping in figs. 3 and 4 
becomes compressed, and the breaking point in fig. 3 is at $x=$0.13 instead of at 0.18,
which would fit better to
the results by Yamada et. al. \cite{yam}. The values of $N(E_F)$
will go down slightly, especially for small $x$.  This
shows that details of the real band structure can be important
for the quantitative results.
 Ab-initio bands in undoped HgBa$_2$CuO$_4$ \cite{tj1} suggest that the
FS retracts towards the diagonal.

The relation between doping and wavelength from the (1-D) band calculations for long supercells appears
to be very exact. For instance, a doubling of a cell will result precisely to a factor 2 in the doping. 
This is because practically all $E_k$ (near $E_F$) 
in the compressed 1-D BZ become gapped. Many states in the interior
of the zone in the 2-D NFE model are not concerned by gaps. All states contribute to the DOS 
and the energies of local gaps do not correspond to the pseudogap in the total DOS.
One could return to the more 'exact' relation
if the 2-D sheet of the NFE band behaved very rigidly with similar gaps everywhere. 

The T-dependent Fermi-Dirac occupation leads to a quenching of the spin wave by the feedback of
the spin density on to the potential. The quenching temperature defines $T^*$, which 
in a 1-D model is about $\frac{1}{4}\Delta/k_B$, where $\Delta \approx V_Q^{sf}$ is the gap at $T$=0 \cite{tj5}.
Figure 4 shows $T^*$ and $N(E_F)$ as function of the doping from the model.

The change of $\vec{q}$ as function of coupling strength was
not forseen in the 1-D band calculations. If this fact is true, then one can expect
longer wave length for weaker coupling,  near $T^*$ for instance.  
This implies that the gap moves away
from $E_F$, which is less favorable for the stability of the wave, and the
process of quenching at $T^*$ may accelerate. The same mechanism predicts isotope
shifts on the q-vectors of magnetic fluctuations, since a heavier mass is expected
to decrease $V_Q^{sf}$ through smaller $u$ \cite{tj5}. 

The 1-D band calculations show that a phonon and a spin wave, which along [1,0,0] differ
by a factor of two in wave length, tend to open a gap at the same energy.  However, the spacial
shape of the potential perturbation caused by a phonon and a spin wave is different, 
(this is less clear from the NFE-model)
and regarding superconductivity
it is not expected that the $V_Q$'s of the two waves work together. The spin wave,
and equal-spin pairing (ESP),
is probably most important for superconductivity, 
since $V_Q^{sf}$ can become very large, as when a favorable phonon displacement is assisting. 
 These arguments suggest ESP as a mechanism
for superconductivity, but it has to compete with the pseudogap
which removes states and DOS near $E_F$, more so at low doping, see figure 4. This scenario
is corroborated by recent femtosecond spectroscopic measurements on cuprates
showing competing order from something like a pseudogap within the superconducting gap \cite{chia}.
The coupling for spin-fluctuations
decreases towards the over-doped side,
as is reflected by the decreasing $V_Q^{sf}$-values in Table I. Also the coupling to
phonons disappears at too short wave lengths, since no spin wave shorter than 4$a_0$ can 
co-exist with a "half-breathing"
phonon. The coupling parameter $\lambda \sim N \cdot V^2$ 
and $T_C$ will therefore vanish at the extreme dopings, as can be deduced from figure 4 by the
low DOS for $x \rightarrow 0$ and the small $T^*$  (which is proportional to $V_Q^{sf}$) 
at large doping. 
The present results are not sufficiently complete for an evaluation of $T_C$ through
a BCS-like formula, but qualitatively it is expected that the limits of a "$T_C$-dome" are
shaped by the lines for $T^*$ and $N(E_F)$, as in figure 4. In order to increase $T_C$ on the
under-doped side one should increase $N(E_F)$. One possiblity is to make the 2-D sheet of 
the band less rigid in
order to restrict the gaps to the exterior of the BZ, perhaps by pressure.

Many typical high-$T_C$ features, such as
FS-arcs, pseudogaps, $T^*$, and the $q(x)$-dependence, can be described qualitatively by the NFE simulations.
The total $V_Q$ determines the size of the pseudogap, but $V_Q^{sf}$ disappears
above $T^*$ and the same gap cannot be supported by $V_Q^{ph}$ alone. The smaller $V_Q^{ph}$
makes a weak dip in the DOS at lower energy, away from $E_F$ and away from
optimal doping. 
It is suggested that superconductivity is caused
by ESP, and $V_Q^{sf}$, which leads to confinement of $T_C$ between two limiting dopings.
  The parameters are based on previous band calculations for phonons
and spin waves in doped systems, and the values in Table I are the most probable $V_Q$
to use in the 2D-NFE model. Still, it is unavoidable that some of the estimations are 
very uncertain. But finally, the extreme simplicity of 2-D NFE model makes it a toy 
model, where other solutions of the parameter space can be tested.

I am grateful to B. Barbiellini and C. Berthod for various discussions.

\end{document}